\documentclass[12pt,a4paper]{article}
\usepackage{a4wide}
\usepackage{amsfonts}
\begin{document}
{\renewcommand{\thefootnote}{\fnsymbol{footnote}}
\hfill  PITHA -- 99/33\\
\medskip
\hfill gr--qc/9910104\\
\medskip
\begin{center}
{\LARGE  Loop Quantum Cosmology II: Volume Operators}\\
\vspace{1.5em}
Martin Bojowald\footnote{e-mail address:
{\tt bojowald@physik.rwth-aachen.de}}\\
Institute for Theoretical Physics, RWTH Aachen\\
D--52056 Aachen, Germany\\
\vspace{1.5em}
\end{center}
}

\setcounter{footnote}{0}

\newtheorem{theo}{Theorem}
\newtheorem{lemma}{Lemma}

\newcommand{\proofend}{\raisebox{1.3mm}{\fbox{\begin{minipage}[b][0cm][b]{0cm}
\end{minipage}}}}
\newenvironment{proof}{\noindent{\it Proof:} }{\mbox{}\hfill 
  \proofend\\\mbox{}}

\newcommand{\Haux}{{\cal H}_{\mbox{\small\rm aux}}}
\newcommand{\ULRS}{{\cal U}^{[\lambda]}_{\mathrm{LRS}}}
\newcommand{\Uiso}{{\cal U}^{[\lambda]}_{\mathrm{iso}}}

\newcommand{\md}{{\mathrm{d}}}
\newcommand{\Aut}{\mathop{\mathrm{Aut}}}
\newcommand{\Ad}{\mathop{\mathrm{Ad}}\nolimits}
\newcommand{\ad}{\mathop{\mathrm{ad}}\nolimits}
\newcommand{\Hom}{\mathop{\mathrm{Hom}}}
\newcommand{\Ima}{\mathop{\mathrm{Im}}}
\newcommand{\id}{\mathop{\mathrm{id}}}
\newcommand{\diag}{\mathop{\mathrm{diag}}}
\newcommand{\Kern}{\mathop{\mathrm{ker}}}
\newcommand{\tr}{\mathop{\mathrm{tr}}}
\newcommand{\sgn}{\mathop{\mathrm{sgn}}}
\newcommand{\semidir}{\mathrel{\mathrm{\times\mkern-3.3mu\protect%
\rule[0.04ex]{0.04em}{1.05ex}\mkern3.3mu\mbox{}}}}
\newcommand{\Vol}{\mathop{\mathrm{Vol}}\nolimits}
\newcommand{\dive}{\mathop{\mathrm{div}}}
\newcommand{\Diff}{\mathop{\mathrm{Diff}}\nolimits}

\newcommand*{\R}{{\mathbb R}}
\newcommand*{\N}{{\mathbb N}}
\newcommand*{\Z}{{\mathbb Z}}
\newcommand*{\Q}{{\mathbb Q}}
\newcommand*{\C}{{\mathbb C}}

\begin{abstract}
  Volume operators measuring the total volume of space in a loop
  quantum theory of cosmological models are constructed. In the case
  of models with rotational symmetry an investigation of the Higgs
  constraint imposed on the reduced connection variables is necessary,
  a complete solution of which is given for isotropic models; in this
  case the volume spectrum can be calculated explicitly. It is
  observed that the stronger the symmetry conditions are the smaller
  is the volume spectrum, which can be interpreted as level splitting
  due to broken symmetries. Some implications for quantum cosmology
  are presented.
\end{abstract}

\section{Introduction}

In this second part we continue the investigation of quantum symmetry
reduction for cosmological models started in the first part
\cite{cosmoI}. There we presented kinematical properties: The general
framework of quantum symmetry reduction \cite{SymmRed} was specialized
to transitive symmetry groups by means of which homogeneous models can
be described. Furthermore, we quantized and solved the Gau\ss\ and
diffeomorphism constraints for all these models. The treated models
are, in order of increasing symmetry, Bianchi class A models
(anisotropic), locally rotationally symmetric (LRS, \cite{Ellis})
models, and isotropic models. For models with a nontrivial isotropy
subgroup, LRS and isotropic models here, there is a further
kinematical constraint, the Higgs constraint, which emerges in the
context of symmetry reduction. A complete solution of this constraint
has not been given, neither in the general framework of reference
\cite{SymmRed} nor in the special cases of reference \cite{cosmoI}. In
the present paper we deal with this constraint in detail for isotropic
models, in which case we present a complete solution, thereby
determining all kinematical states. This task is complicated by the
fact that the quantum configuration space is not a group implying that
kinematical quantum states are not given by ordinary spin networks.
For LRS models the treatment is analogous. All these models serve as
examples for a solution of the Higgs constraint in the general
framework.

Moreover, we will use here these kinematical Hilbert spaces to
quantize operators measuring the total volume of space and to
investigate their spectra. As a first application of quantum
symmetry reduction it has been observed in reference \cite{AreaOp}
that the area spectrum in spherically symmetric sectors of loop
quantum gravity is only a small subset of the full spectrum. The huge
spectrum of the non-symmetric operator was interpreted as consequence
of a level splitting caused by broken spherical symmetry. The same
phenomenon will be observed here for the volume spectrum. Its
phenomenology is richer in this case, because we can relax symmetry
conditions in steps: Starting with isotropic models we can first
proceed to locally rotationally symmetric models with only one axis,
followed by anisotropic but still homogeneous Bianchi models, and
finally break the symmetry completely to reach the full theory. In
each step a part of the maximal symmetry group is broken,
and in each step the volume spectrum is enlarged by new eigenvalues
and possibly a shift in the old ones. Whereas the spectrum of the full
theory is very complex --~the eigenvalues can be given explicitly only
in simple cases \cite{Vol}~-- the volume spectrum for isotropic models
can be calculated explicitly. In view of the important role
which the volume operator plays also for dynamics \cite{QSDI} this
simplification of its spectrum, besides a geometric simplification of
isotropic spin network states, shows that isotropic models can be good
test models to understand the Wheeler--DeWitt equation in quantum
gravity \cite{cosmoIII}. Contrary to most former treatments of quantum
cosmological models this Wheeler--DeWitt equation of loop quantum
cosmology is a discrete equation, not a differential equation, e.g.\ 
in the scale factor of the universe, as in minisuperspace
quantizations. This is a manifestation of the discrete structure of
space revealed in loop quantum gravity.

In the next section we will recall the kinematical properties of the
models treated in reference \cite{cosmoI}. In Section~\ref{s:bianchi}
the volume of Bianchi class A models will be quantized and compared
with the volume operator of the full theory. In this case the isotropy
subgroup is trivial, and therefore the Higgs constraint is empty. But
in case of LRS and isotropic models we have to solve the Higgs
constraint in quantum theory, which will be done in
Section~\ref{s:iso} for isotropic models in detail. The results are
used to quantize the volume operator and to calculate its spectrum.
For LRS models the treatment will mainly be analogous to isotropic
models but not given completely in this paper. Finally, in
Section~\ref{s:applic} we present some applications, e.g.\ 
construction of weave states and some cosmological implications.

\section{Bianchi, LRS and Isotropic Models}

The setting for implementing a (quantum) symmetry reduction is a
symmetry group $S$ acting on a principal fiber bundle
$P(\Sigma,G,\pi)$ over the space manifold $\Sigma$ which is here
assumed to be compact (this is only for ease of
presentation, otherwise the framework has to be adapted
appropriately). The structure group is $G=SU(2)$ for gravity in the
real Ashtekar formulation \cite{AshVar,AshVarReell}. A classical
symmetry reduction can be done, in the most general framework, by
using the classification of invariant connections \cite{KobNom}, which
shows that for a transitive symmetry group each invariant connection
can be expressed by some scalar fields (collectively called Higgs
field) subject to a Higgs constraint. This constraint is empty for a
free action of the group $S$, and depends on a homomorphism
$\lambda\colon F\to G$ (more precisely, its conjugacy class) if the
isotropy subgroup $F$ (for a fixed but arbitrary base point $x_0$ in
$\Sigma$) of $S$ is nontrivial. The space manifold $\Sigma$ can be
identified with $S/F$ or an appropriate compactification thereof.
This framework is specialized to cosmological models in reference
\cite{cosmoI}, and its results will now be recalled briefly.

The models of interest are Bianchi class A models with a freely acting
symmetry group, i.e.\ $F=\{1\}$, and LRS and isotropic models, for
which the symmetry group can be written as a semidirect product
$S=N\semidir_{\rho}F$ with the translation subgroup $N$ and the
isotropy subgroup $F$. The representation $\rho\colon F\to\Aut N$
describes how the isotropy subgroup acts on the tangent space ${\cal
  L}N$ of a point in $\Sigma$. For LRS models we have $F=U(1)$ and for
isotropic models $F=SU(2)$, $\rho$ acting in both cases by rotations.
An invariant connection can always be written as
$A=\phi^i_I\omega^I\tau_i$, where $\tau_j=-\frac{i}{2}\sigma_j$ (using
the Pauli matrices $\sigma_j$) are generators of $G=SU(2)$ and
$\omega^I$ left invariant one-forms on $N$ (for Bianchi models we set
$N:=S$). The components $\phi^i_I$ of a linear map $\phi\colon{\cal
  L}N\to{\cal L}G$ are collectively denoted as Higgs field.  For
Bianchi models these components are unrestricted, whereas they are
restricted by the Higgs constraint to be of the form
\[
  \phi^i_1=2^{-\frac{1}{2}}(a\Lambda^i_1+b\Lambda^i_2)\quad,\quad
  \phi_2^i=2^{-\frac{1}{2}}(-b\Lambda^i_1+a\Lambda^i_2)\quad,\quad
  \phi^i_3=c\Lambda^i_3
\]
for LRS models and $\phi^i_I=c\Lambda^i_I$ for isotropic models,
respectively, with a fixed but arbitrary dreibein $\Lambda$.  The
dreibein $\Lambda$ depends on the homomorphism $\lambda\colon F\to G$
chosen in its conjugacy class. Fixing such a homomorphism and,
therefore, $\Lambda$ amounts to a partial gauge fixing which will be
undone in the quantum theory. Without gauge fixing $\Lambda$ is
arbitrary but pure gauge.

The momenta conjugate to the connections above are invariant (with
respect to the $S$-action) density-weighted dreibeine given by
$E^a_i=\sqrt{g_0}\,p^I_iX^a_I$ in terms of left invariant vector
fields $X_I$ obeying $\omega^I(X_J)=\delta^I_J$. For Bianchi models
the $p^I_i$ are arbitrary and conjugate to $\phi^i_I$,
whereas they are restricted to be of the form
\[
  p^1_i=2^{-\frac{1}{2}}(p_a\Lambda^i_1+p_b\Lambda^i_2)\quad,\quad
  p_i^2=2^{-\frac{1}{2}}(-p_b\Lambda^i_1+p_a\Lambda^i_2)\quad,\quad
  p_i^3=p_c\Lambda^i_3
\]
for LRS, and $p^I_i=p\Lambda^I_i$ for isotropic models. The density
weight is provided by the determinant $g_0$ of the left invariant
metric on $\Sigma$ defined by $\omega^1\wedge\omega^2\wedge\omega^3=
\sqrt{g_0}\,\md^3x$.

From these momenta expressions for the volume are built as follows:
\begin{eqnarray}\label{volbianchi}
 V & = & \int_{\Sigma}\md^3x \sqrt{{\textstyle\frac{1}{6}}\left|
  \epsilon^{ijk}\epsilon_{abc} E^a_i E^b_j E^c_k\right|}=
  \int_{\Sigma}\md^3x
 \sqrt{{\textstyle\frac{1}{6}} g_0^{\frac{3}{2}}
   \left|\epsilon^{ijk}\epsilon_{abc} p^I_i
   p^J_j p^K_k X^a_I X^b_J X^c_K\right|}\nonumber\\
 & = & \int_{\Sigma}\md^3x \sqrt{{\textstyle\frac{1}{6}} g_0^{\frac{3}{2}}
   \left|\epsilon^{ijk} \epsilon_{IJK} p^I_i p^J_j p^K_k 
   \det(X^a_L)\right|}=
   V_0 \sqrt{{\textstyle\frac{1}{6}} \left|\epsilon^{ijk}\epsilon_{IJK}
   p^I_i p^J_j p^K_k\right|}
\end{eqnarray}
using $\det X^a_I=(\det\omega^I_a)^{-1}=g_0^{-\frac{1}{2}}$ and
$V_0:=\int_{\Sigma}\md^3x\sqrt{g_0}$. For LRS
models this leads to
\begin{equation}\label{volLRS}
 V=V_0\sqrt{{\textstyle\frac{1}{2}}(p_a^2+p_b^2)|p_c|}\,,
\end{equation}
and for isotropic models to
\begin{equation}\label{volisotropic}
 V=V_0|p|^{\frac{3}{2}}\,.
\end{equation}

The basic ingredient for a quantum symmetry reduction \cite{SymmRed}
is a pull back map from the space of functions on the space of
connections on $\Sigma$, which is the auxiliary Hilbert space of the
full theory, to a space of functions on the space of fields
classifying invariant connections, i.e.\ to functions on spaces of
Higgs fields. In quantum theory one uses certain extensions of the
spaces of connections and Higgs fields, which can in the case of Higgs
fields best be described in terms of point holonomies
\cite{FermionHiggs}. For Bianchi models there are three Higgs `fields'
$\phi^i_I\tau_i$, $1\leq I\leq 3$ in a single point $x=0$ (strictly
speaking, they are no longer fields in a homogeneous context) leading
to three point holonomies. These can be extended to ordinary
holonomies by reintroducing an auxiliary manifold $\overline{S/F}$ in
which the point holonomies are written as holonomies associated with
three edges $e_I$ parallel to the invariant vector fields $X_I$ (the
auxiliary manifold should be compactified such that the edges are
closed curves). Then the auxiliary Hilbert space
$\Haux=L^2(SU(2)^3,\md\mu_H^3)$ ($\md\mu_H$ is the Haar measure on
$SU(2)$) is spanned by spin networks associated with graphs containing
three closed edges meeting in the $6$-vertex $x_0$ which is the base
point chosen in $\Sigma$. The Gau\ss\ constraint enforces gauge
invariance of those spin networks, i.e.\ the six edge representations
(each edge is incoming and outgoing) are to be contracted to the
trivial representation in $x_0$. This auxiliary Hilbert space
illustrates the reduction of degrees of freedom to finitely many ones
by the symmetry reduction.

Up to now all Bianchi class A models are presented on the same
auxiliary Hilbert space. Differences are introduced already at the
kinematical level by the diffeomorphism constraint: It enforces
invariance under inner automorphisms acting on $\overline{S/F}$, which
can be interpreted as independence of choosing the base point $x_0$
\cite{cosmoI}. Inner automorphisms are certainly sensitive to the
algebraic structure of the symmetry group. E.g., for the Bianchi I
model with $S=\R^3$ they are all trivial, and therefore the
diffeomorphism constraint is empty. For Bianchi IX with $S=SU(2)$,
however, the group of inner automorphisms is isomorphic to $SO(3)$
acting on $\overline{S/F}=S^3$ by rotations. Therefore, all rotated spin
networks are equivalent leading after group averaging to linear
combinations of spin network states which are invariant under
permutation of the edge spins. This reduces the number of allowed spin
networks and affects the volume spectrum (but only slightly), as we
will see below. These two Bianchi class A models are most interesting
for our purposes, because they can be reduced further to isotropic
models.

In models with a nontrivial isotropy subgroup, LRS and isotropic
models, the situation is more complicated. Here we have the Higgs
constraint, which is easy to solve classically, but which implies
that in the quantum theory we will no longer have functions on a group
($SU(2)^3$ above) but on a certain union of conjugacy classes which
is not a subgroup. For use in quantum theory the Higgs constraint can
advantageously be written as
\begin{equation}\label{Higgs}
 h(\rho(f)(e_I))=\exp(\lambda(f)) h(e_I)\exp(-\lambda(f))
\end{equation}
where $\rho$ is the action of the isotropy subgroup $F$ on
$N\cong S/F$ (or $\overline{S/F}$ after compactification), and
$\lambda\colon F\to G$ the homomorphism introduced above. In the
rotationally symmetric models $\lambda$ will embed $F=U(1)$ as a
subgroup of $G=SU(2)$ for LRS models, or be the identity for isotropic
models. With $h(e_I)$ we denote the holonomy associated to the edge
$e_I$ in the auxiliary manifold for a fixed Higgs field.  The Higgs
constraint is thus interpreted geometrically as saying that holonomies
to edges which are rotated by elements of $F$ are gauge equivalent.
Therefore, one would expect that this constraint can be solved by
using a special class of spin networks: For LRS models two of the
three holonomies of homogeneous spin networks
are gauge equivalent, and for isotropic models all three holonomies.
Spin networks for LRS models should then consist of only two edges, an
axial one and a transversal one representing the two equivalent edges,
and for isotropic models of only one edge. However, this consideration
takes into account only the edges, not the vertex contractor which is
an additional labeling. A reduction of this contractor is not obvious
from the constraint (\ref{Higgs}). Indeed we will see in
Section~\ref{s:iso} that there is an insertion in the single edge of
isotropic spin networks, which can be seen as a remnant of the vertex
contractor. This insertion enables a non-vanishing volume, which shows
its necessity from another viewpoint because the volume operators
need vertices (more than $3$-valent for gauge invariant vertices) to
act on non-trivially.

These isotropic spin networks will be found by studying functions on
the quantum configuration space
\begin{equation}\label{Uiso}
  {\cal U}^{[\lambda]}_{\mathrm{iso}}=\left\{\left(\exp
      (c\Lambda_1^i\tau_i),\exp
  (c\Lambda_2^i\tau_i), \exp (c\Lambda^i_3\tau_i)\right)\right\}
\end{equation}
which is obtained by exponentiating the classical solution space of
the Higgs constraint. It is a union of conjugacy classes in $SU(2)^3$
labeled by $c$, and the gauge group $G=SU(2)$ acts on it by diagonal
conjugation:
$g(h_1,h_2,h_3)g^{-1}=(gh_1g^{-1},gh_2g^{-1},gh_3g^{-1})$. This shows
that the dreibein $\Lambda$ is pure gauge, but it is needed to undo
gauge fixing. Relaxing the gauge fixing is necessary to be able to
further on use $SU(2)$-spin networks and point holonomies. The fact
that ${\cal U}^{[\lambda]}_{\mathrm{iso}}$ is no longer a group
implies that quantum states, i.e.\ functions thereon, are no longer
ordinary spin networks.  These are usually obtained by making use of
the Peter--Weyl theorem which determines all functions on a group. It
can now no longer be used, and we will have to determine all gauge
invariant functions by hand.  This leads to the possibility of
insertions mentioned above, which do not appear in ordinary spin
networks.

\section{Volume Operator for Bianchi Class A Models}
\label{s:bianchi}

Acting on functions in $\Haux$ the momentum operators are represented as
\begin{equation}\label{quantp}
  \hat{p}^I_i=\frac{\iota' l_P^2}{2}\left(J_i^{(R)}(h_I)+
    J_i^{(L)}(h_I)\right)
\end{equation}
where $J^{(R)}(h_I):=-iX_i^{(R)}(h_I)$ and
$J^{(L)}(h_I):=-iX_i^{(L)}(h_I)$ are right and left invariant
selfadjoint angular momentum operators defined via the right and left
invariant vector fields acting on the copy of $SU(2)$ associated with
the edge $e_I$. Furthermore, $\iota$ is the Immirzi parameter,
$\iota':=\iota V_0$, and $l_P$ the Planck length. The appearance of
both right and left invariant vector fields is due to the fact that
each of the edges $e_I$ is both incoming and outgoing in the vertex
$x_0$.

These operators can now be inserted in the classical expression
(\ref{volbianchi}) to obtain the volume operator
\begin{equation}
 \hat{V} = V_0\sqrt{{\textstyle\frac{1}{6}} \left|\epsilon^{ijk}
   \epsilon_{IJK} \hat{p}^I_i\hat{p}^J_j\hat{p}^K_k\right|}
   = V_0 (\iota' l_P^2)^{\frac{3}{2}}
 \sqrt{|\hat{q}|}\label{volbianchiquant}
\end{equation}
with the operator
\begin{eqnarray}\label{volO}
  \hat{q}\!\!\! & = & \!\!\!{\textstyle\frac{1}{48}}\sum_{I,J,K=1}^3
   \epsilon^{ijk}\epsilon_{IJK}
   \left(J_i^{(R)}(h_I)+J_i^{(L)}(h_I)\right)\!\!
   \left(J_j^{(R)}(h_J)+J_j^{(L)}(h_J)\right)\!\!
   \left(J_k^{(R)}(h_K)+J_k^{(L)}(h_K)\right)\nonumber\\
  & = &\!\!\! {\textstyle\frac{1}{8}}\epsilon^{ijk}
  \left(J_i^{(R)}(h_1)+J_i^{(L)}(h_1)\right)
   \left(J_j^{(R)}(h_2)+J_j^{(L)}(h_2)\right)
   \left(J_k^{(R)}(h_3)+J_k^{(L)}(h_3)\right)
\end{eqnarray}

It is to be compared with the contribution to the volume operator of
the full theory \cite{Vol2} in a single vertex. Here we have the
$6$-vertex with (after cutting each of the closed edges $e_I$ in two
pieces) three incoming and three outgoing edges, each edge
contributing either by a left or right invariant vector field. If we
expand the product in $\hat{q}$ of the three terms containing the
derivative operators, we obtain a sum of terms each being a gauge
invariant product of angular momentum operators of the form
$\epsilon^{ijk} \epsilon_{IJK} J_i^{(R/L)}(h_I) J_j^{(R/L)}(h_J)
J_k^{(R/L)}(h_K)$. These correspond to all non-planar sets of three
edges incident in $x_0$, and the factor $\epsilon_{IJK}$ introduces
the correct sign for the dreibein of the associated three edges. Thus,
we see that the operator in the single vertex here equals exactly a
vertex contribution of the full operator. The scale factor $V_0$ is
different from the full operator (and arbitrary, for we could choose
another metric $g_0$), but note that the operator of reference
\cite{Vol2} also contains an arbitrary scale factor, called $\kappa_0$
there, as a relic of the regularization. The only important difference
between the symmetric and the non-symmetric operators is the missing
vertex sum for the non-symmetric one. This is analogous to the area
operator in spherically symmetric sectors \cite{AreaOp}. Note that we
are lead naturally to the operator of reference \cite{Vol2} by using
the quantization (\ref{quantp}) of the $p^I_i$ which is forced on us
by the general treatment of point holonomies. The alternative operator
of reference \cite{AreaVol}, however, cannot be obtained in the
present context (it contains absolute squares for each triple product
of angular momentum operators not just for the sum; for a comparison
of the operators see reference \cite{VolVol}).

Although the vertices appearing here are at most $6$-valent leading to
a slight simplification of the volume operator, it is impossible to
calculate all eigenvalues explicitly. The vertices appearing here can,
however, all be found also in a lattice formulation of loop quantum
gravity \cite{Loll:vol}. Hence, the techniques developed in reference
\cite{Loll:Simply} by using the octagonal group can be employed to
determine the volume spectrum of Bianchi models.

The operator (\ref{volbianchiquant}) is valid for all Bianchi class A
models irrespective of the particular type. However, the volume
spectrum depends on the type, because the diffeomorphism constraint
selects special linear combinations of spin networks. The greater the
group of inner automorphisms of $S$ the smaller is the volume spectrum
(see the remarks in the preceding section, and reference \cite{cosmoI}
for more details). These are only minor changes of the spectra as
compared to the changes introduced by symmetry conditions, which we
will study now.

\section{Solving the Higgs Constraint and Volume Operator for
  Isotropic Models}
\label{s:iso}

As a consequence of the Higgs constraint (\ref{Higgs}) not all three
holonomies are independent if there is a nontrivial isotropy subgroup
leading to the following relations between invariant vector fields:
$h_I=gh_Jg^{-1}$ implies
\begin{eqnarray}
 X_i^{(R)}(h_I) & = & \tr\left[(\tau_ih_I)^T
    \frac{\partial}{\partial h_I}\right]= \tr\left[
   (\tau_igh_Jg^{-1})^T g^{-T}\frac{\partial}{\partial h_J}
    g^T\right]\nonumber\\
 & = & \tr\left[ (g^{-1}\tau_igh_J)^T\frac{\partial}{\partial h_J}
  \right]= \Ad(g^{-1})_{ij}
 X_j^{(R)}(h_J)
\end{eqnarray}
with the matrix elements $\Ad(g)_{ij}$ defined by
$g\tau_ig^{-1}=:\Ad(g)_{ij}\tau^j$. This implies
\begin{equation}
  X_i^{(R)}(h_I)\tau^i=\Ad(g^{-1})_{ij}X_j^{(R)}(h_J)\tau^i=
  gX_j^{(R)}(h_J)\tau^jg^{-1}
\end{equation}
and analogously for $X_i^{(L)}$. This equation can be used to derive
the volume operators for LRS and isotropic models from the operator
(\ref{volbianchiquant}) for Bianchi models.

The essential ingredient of equation (\ref{volO}) is
$\epsilon^{ijk}J^1_iJ^2_jJ^3_k$ (we define
\[
  J^I_i:={\textstyle\frac{1}{2}}\left(J_i^{(R)}(h_I)+J_i^{(L)}(h_I)\right)
\]
and later
$J^I:=J^I_i\tau^i$ for ease of
notation), which can be written as
$-4\tr(J^1_i\tau^iJ^2_j\tau^jJ^3_k\tau^k)$. For LRS models we insert
\[
  h_2=\exp\left({\textstyle\frac{\pi}{2}}\Lambda_3^i\tau_i\right) h_1
  \exp\left(-{\textstyle\frac{\pi}{2}}\Lambda_3^i\tau_i\right)
\]
yielding
\begin{eqnarray}
  -4\tr(J^1J^2J^3) & = &
  -4\tr\left[J^1_i\tau^i\exp\left({\textstyle\frac{\pi}{2}}
      \Lambda_3^l\tau_l\right) J^1_j\tau^j
  \exp\left(-{\textstyle\frac{\pi}{2}}\Lambda_3^m\tau_m\right)
  J^3_k\tau^k\right]\nonumber\\
  & = & -4J^1_iJ^1_jJ^3_k \tr\left[\tau^i(\epsilon_{jml}\tau^m\Lambda_3^l+
  \Lambda_3^j\Lambda_3^m\tau_m) \tau^k\right]\nonumber\\
  & = & -J^1_iJ^1_jJ^3_k(\Lambda_3^j\delta^{ik}-\Lambda_3^k\delta_{ij}+
  \Lambda_3^j\Lambda_3^l\epsilon_{ikl})\nonumber\\
  & = & -\Lambda_3^i J^1_i J^1_j J^{3j}+\Lambda_3^i J^3_i J^1_j
  J^{2j}- \Lambda_3^i J^1_i
  \Lambda_3^l\epsilon_{ikl}J^{1i}J^{3k}\,.\label{JLRS}
\end{eqnarray}
Analogously we obtain in case of isotropic models
\begin{eqnarray}
  -4\tr(J^1J^2J^3) & = & -4\tr\left[\exp\left({\textstyle\frac{\pi}{2}}
      \Lambda_2^l\tau_l\right)
  J_i^3\tau_i
  \exp\left(-{\textstyle\frac{\pi}{2}}\Lambda_2^m\tau_m\right)
  \right.\nonumber\\
  & & \qquad
  \left.\exp\left(-{\textstyle\frac{\pi}{2}}
      \Lambda_1^n\tau_n\right)J_j^3\tau_j
  \exp\left({\textstyle\frac{\pi}{2}}\Lambda_1^o\tau_o\right)
  J_k^3\tau_k\right]\nonumber\\
  & = & -4J_i^3J_j^3J_k^3 \tr\left[(\epsilon_{pli}\tau^p\Lambda_2^l+
  \Lambda_2^i\Lambda_2^m\tau_m)(-\epsilon_{qnj}\tau^q\Lambda_1^n+
  \Lambda_1^j\Lambda_1^o\tau_o)\tau_k\right]\nonumber\\
  & = & J_i^3J_j^3J_k^3\left(\epsilon_{jli}\Lambda_2^l\Lambda_1^k-
  \epsilon_{nli}\Lambda_2^l\Lambda_1^n\delta_{jk}-
  \delta_{ik}\Lambda_2^l\Lambda_1^j\Lambda_{1l}\right.\nonumber\\
  & & \left.+\Lambda_2^k\Lambda_1^j\Lambda_1^i+
  \Lambda_2^i\Lambda_2^j\Lambda^k_1-
  \delta_{jk}\Lambda_2^i\Lambda_2^m\Lambda_{1m}-
  \epsilon_{mok}\Lambda_2^i\Lambda_2^m\Lambda_1^j\Lambda_1^o
   \right)\nonumber\\
  & = & J_i^3J_j^3J_k^3(-\delta_{jk}\Lambda_3^i)= -\Lambda_3^i J_i^3
  (J^3)^2\label{Jiso}
\end{eqnarray}
using $\Lambda_I^i\Lambda_{Ji}=\delta_{IJ}$ and $\Lambda_I^i
J_i^3\propto\delta_{I3}$ (the last relation will be established in
Subsection \ref{s:deriv}).

We see in these preliminary expressions that only derivative operators
for the independent holonomies $h_1$, $h_3$ for LRS and $h_3$ for
isotropic models appear. But they contain operators of the form
$\Lambda_3^i J^I_i$ whose action we do not know yet. Note that these
operators are gauge invariant: by definition $\Lambda_3^i\tau_i$
transforms by conjugation under a gauge transformation. Therefore,
$\Lambda_3^i J^I_i= -2\tr(\Lambda_3^i\tau_iJ_j^I\tau^j)$ is gauge
invariant. We have here already undone the partial gauge fixing,
meaning that $\Lambda_I^i$ is not a fixed dreibein, but transforms
under gauge transformations which change the gauge fixing. The unit
vectors $\Lambda_I$ have to be regarded as functions $\Lambda_I\colon
SU(2)^3\to S^2,\Lambda_I(g_1,g_2,g_3)=L(g_I)$ for $g_I\not=1$ from the
classical configuration space of Bianchi models to the unit sphere
embedded in the Lie algebra of $SU(2)$. The function $L\colon
SU(2)\backslash\{1\}\to S^2=\{X\in {\cal L}SU(2):X^iX_i=1\}$ is
defined for $g\not=1$ as $L(g):=\left[(\log g)^i(\log
g)_i\right]^{-\frac{1}{2}}\log g$ using the matrix logarithm which can
be made unique by fixing a branch, e.g.\ by demanding that $\log
g\in{\cal L}SU(2)$ has minimum Cartan--Killing norm. As a consequence,
the operators $\Lambda_3^i J^3_i$ are not symmetric, although
$\Lambda_3^i$ are real functions and the $J^I_i$ are self-adjoint:
$(\Lambda_3^i J^3_i)^*=J^3_i\Lambda_3^i\not= \Lambda_3^i J^3_i$
because of $[\Lambda_3^i,J^3_j]\not=0$. The commutator is a
complicated function on $SU(2)$ due to the logarithms in the
definition of $\Lambda_3$. We also see that there are factor ordering
ambiguities in the expressions (\ref{JLRS}) and (\ref{Jiso}) which we
ignored above.

To understand the action of the operator in equation (\ref{Jiso}) we
have to gain more knowledge about the quantum states of isotropic
models. To compute the complete spectrum of the volume operator we
have to know all these states.

\subsection{Quantum States for LRS and Isotropic Models}

In the course of quantum symmetry reduction quantum states are defined
as functions on the spaces (\ref{Uiso}) for isotropic models and
\begin{equation}\label{ULRS}
  {\cal U}^{[\lambda]}_{\mathrm{LRS}}=\left\{\left(\exp
      (2^{-\frac{1}{2}}(a\Lambda_1^i+ 
  b\Lambda_2^i)\tau_i),\exp(2^{-\frac{1}{2}}(-b\Lambda_1^i+
  a\Lambda_2^i)\tau_i), \exp (c\Lambda^i_3\tau_i)\right)\right\}
\end{equation}
for LRS models. These spaces are obtained by exponentiating the
solution spaces of the classical Higgs constraint, and their elements
solve the Higgs constraint in the form (\ref{Higgs}). They are
parameterized by the parameters $\Lambda_I^i$, which are pure gauge
but arbitrary after relaxing the gauge fixing, and $a$, $b$, $c$ for
LRS models and $c$ for isotropic models, respectively. Furthermore,
they are submanifolds of the configuration space $SU(2)^3$ of Bianchi
models, but not subgroups. All functions on them can be generated by
pull backs of spin network functions on $SU(2)^3$, but not all of
these pull backs will be independent. Pull backs of gauge invariant
spin networks only depend on $c$ for isotropic models, and on
$A:=\sqrt{a^2+b^2}$ and $c$ for LRS models; they are automatically
gauge invariant on the reduced configuration spaces. This implies that
the parameterizations are highly redundant: Gauge invariant functions
can be expressed as functions only in one $SU(2)$-element, which we
choose as the third, for ${\cal U}^{[\lambda]}_{\mathrm{iso}}$, and in
the first (choosing this one of the first two) and third element for
${\cal U}^{[\lambda]}_{\mathrm{LRS}}$.  This corresponds to the fact
that equation (\ref{Higgs}) eliminates the holonomies $h_1$ and $h_2$
for isotropic models, and $h_2$ for LRS models. A special class of
such functions is given by spin network functions associated with
graphs containing only one edge $e_3$ for isotropic models, or two
edges $e_1$, $e_3$ for LRS models. But these do not suffice to
generate all gauge invariant functions, neither for LRS nor for
isotropic models.  To show this we use a small lemma, which will also
prove useful when calculating particular spin networks:

\begin{lemma} Let $g:=\exp (A\tau_i)$ and $h:=\exp (B\tau_j)$ with
  $A,B\in\R$, $i\not=j$ be matrices in the fundamental representation of
  $SU(2)$. Then
 \[ 
   gh=hg+h^{-1}g+hg^{-1}-\tr(gh)\,.
 \]
\end{lemma}

\begin{proof}
  This can directly be proved by using $\exp (A\tau_i)=\cos (A/2)+2\sin
  (A/2)\tau_i$.
\end{proof}

By means of this lemma we can express a product of arbitrary factors
$\exp (A_k\tau_i)$ as a sum of terms with at most three factors. To
calculate the gauge invariant trace we then need only $\tr\exp
(A\tau_i)=2\cos (A/2)$, $\tr [\exp (A\tau_i)\exp (B\tau_j)]=2\cos
(A/2)\cos (B/2)$ for $i\not=j$ and
\[
  \tr[\exp (A\tau_1)\exp (B\tau_2)\exp (C\tau_3)]=2\cos
({\textstyle\frac{1}{2}}A)\cos ({\textstyle\frac{1}{2}}B)\cos
({\textstyle\frac{1}{2}}C)-2\sin ({\textstyle\frac{1}{2}}A)\sin
({\textstyle\frac{1}{2}}B)\sin ({\textstyle\frac{1}{2}}C)\,.
\]

We now show that any pull back of a gauge invariant function on
$SU(2)^3$ to a gauge invariant function on $\ULRS$ is invariant under
the reflection $A=\sqrt{a^2+b^2}\mapsto -A$. An overcomplete set of
such functions on $SU(2)^3$ is given by
$\tr(h_1^{n_1}h_2^{n_2}h_3^{n_3}h_1^{n_4}h_2^{n_5}\cdots)$ with an
arbitrary finite number of factors and arbitrary $n_i$. By using the
lemma these functions can be simplified to
$\tr(h_1^{m_1}h_2^{m_2}h_3^{m_3})$ (up to factors of $\cos (A/2)$ or
$\cos (c/2)$). They are gauge invariant, and so we can choose the
gauge $h_1=\exp (A\tau_1)$, $h_2=\exp (A\tau_2)$, $h_3=\exp (c\tau_3)$.
Using a gauge transformation $g=\exp(\pi\tau_3)=2\tau_3$ (in general
$g=2\Lambda_3^i\tau_i$) with $gh_3g^{-1}=h_3$, $gh_1g^{-1}=h_1^{-1}$,
$gh_2g^{-1}=h_2^{-1}$ we see that all gauge invariant functions are
invariant under $h_1\mapsto h_1^{-1}$, $h_2\mapsto h_2^{-1}$,
$h_3\mapsto h_3$ which is equivalent to $A\mapsto -A$, $c\mapsto c$.
Of course, the gauge invariant spin networks with only two edges
$e_1$, $e_3$ are also invariant under this transformation. The key
point is that there is a gauge transformation which fixes $h_3$, which
depends on $c$, but inverts $h_1$ and $h_2$, which depend on $A$.

The situation is different if we are interested in the transformation
$A\mapsto A$, $c\mapsto -c$: There is no gauge transformation fixing
both $h_1$ and $h_2$, but inverting $h_3$. Therefore, gauge invariant
spin networks on the three edges $e_1$, $e_2$, $e_3$ do not need to be
invariant under $A\mapsto A$, $c\mapsto -c$. A counterexample is
provided by a spin network with three edge spins $\frac{1}{2}$ and an
appropriate gauge invariant vertex contractor such that it can be
written as $\tr[\exp (A\tau_1)\exp (A\tau_2)\exp (c\tau_3)]=2\cos^2
(A/2)\cos (c/2)-2\sin^2 (A/2)\sin (c/2)$. In contrast, reduced spin
networks with the two edges $e_1$, $e_3$ give always rise to gauge
invariant functions being invariant under $A\mapsto A$, $c\mapsto -c$,
which can be shown as above by using the gauge transformation
$g'=\exp(\pi\tau_1)=2\tau_1$.

Thus, we see that the obvious candidates for functions on $\ULRS$,
namely spin network functions with two edges, are not sufficient to
generate all gauge invariant functions. This fact can be traced back
to the twisting in equation (\ref{Higgs}) introduced by the
non-trivial gauge transformation on the right hand side. A related
fact is that $\ULRS$ is not a subgroup of $SU(2)^3$, but only a subset
being a union of conjugacy classes. The Peter--Weyl theorem does no
longer apply in this situation; spin network functions with two edges
could only be expected as a sufficient class of functions if the
reduced configuration space would be a subgroup of $SU(2)^3$, e.g.\ 
$SU(2)^2$ which would be obtained in case of a trivial gauge
transformation in equation (\ref{Higgs}). This sector, however, does
not allow nontrivial Higgs fields \cite{cosmoI}.

An analogous discussion applies for the isotropic models: Gauge
invariant spin network functions with one edge are always invariant
under $c\mapsto-c$, but this is not necessarily true for gauge
invariant functions on $\Uiso$.

\subsection{Insertions}

We now have to face the two problems of investigating the operator
$\Lambda_3^i J^3_i$ (determining its action, a symmetric ordering and
selfadjoint extensions) and of determining all quantum states.
Luckily, these two problems are connected, and one problem will
provide the solution for the other. This can easily be seen by
applying $\Lambda_3^i X_i^{(R)}(h_3)$ to a gauge invariant function on
two edges, which, as shown above, can always be written as a linear
combination of terms $\tr(h_1^{m_1}h_3^{m_3})$ being invariant under
$A\mapsto A$, $c\mapsto -c$. Applying the operator
\[
  2\Lambda_3^i X_i^{(R)}(h_3)=
  2(\Lambda_3^i\tau_ih_3)^A_B\frac{\partial}{\partial (h_3)^A_B}
\]
effects a replacement of $h_3$ with
$2\Lambda_3^i\tau_ih_3=\exp(\pi\Lambda_3^i\tau_i)h_3$:
\[
  2\Lambda_3^i X_i^{(R)}(h_3)\tr(h_1^{m_1}h_3^{m_3})=
  \tr[h_1^{m_1}(\exp(\pi\Lambda_3^i\tau_i)h_3)^{m_3}]
\]
and the same for $\Lambda_3^i X_i^{(L)}(h_3)$.  This function is no
longer invariant under $A\mapsto A$, $c\mapsto -c$: The gauge
transformation $g'=\exp(\pi\tau_1)$ used above transforms
$\exp(\pi\Lambda_3^i\tau_i)h_3$ into
$-\exp(\pi\Lambda_3^i\tau_i)h_3^{-1}$, and therefore the above
function changes sign under $A\mapsto A$, $c\mapsto -c$ for $m_3$ odd.

The operator $\Lambda_3^i J_i^3$, which will appear in the volume
operators, does not fix the space of spin network functions with two
edges. Therefore, we have to understand the remaining states in order
to investigate the volume operators. We can visualize them as spin
networks with two closed edges, but with an additional insertion in
its $4$-vertex, which is associated to the holonomy $h_3$. This
symbolizes the insertion of $\exp(\pi\Lambda_3^i\tau_i)$, and it can be
interpreted, together with the $4$-vertex contractor, as a remnant of
the $6$-vertex contractor of a spin network on $SU(2)^3$ after
reduction to local rotational symmetry. The kinematical Hilbert space
thus splits into a subspace of ordinary spin network functions with
two edges, and a subspace of spin network functions with
insertion. The volume operator will fix neither of these subspaces.

An analogous discussion applies to the case of isotropic models. Here,
there remains only one closed edge after reduction, and the insertion
in its $2$-vertex is the only remnant of the $6$-vertex contractor, or
of the $4$-vertex contractor with the insertion for LRS models. In
this case of isotropic models we will demonstrate in the next
subsection that we now have found all quantum states, and develop a
calculus on the solution space of the Higgs constraint enabling us to
deal with operators like $\Lambda_3^i J^3_i$.

\subsection{Kinematical Hilbert Space of Isotropic Models}

Before discussing in more detail quantum states of isotropic models,
we determine a measure on the space $\Uiso$, which can be regarded
as the space of generalized isotropic connections. This measure will
be derived from the Ashtekar--Lewandowski measure along the lines of
quantum symmetry reduction.

Reducing first to homogeneous connections, the Ashtekar--Lewandowski
measure of the full theory is reduced to the finite-dimensional
measure $\md\mu_H^3$ on $SU(2)^3$. Here we parameterize $SU(2)$ as
$g=\exp (c\Lambda_3^i\tau_i)\in SU(2)$ with $\Lambda_3\in S^2$
leading to the normalized Haar measure
\[
  \mu_H(f)=\frac{1}{4\pi^2}\int_0^{4\pi}\int_0^{\pi}\int_0^{\pi}
  f(c,\Lambda_3^i)\sin^2{\textstyle\frac{c}{2}}\,\md^2\Lambda_3\md c
\]
with the solid angle measure $\md^2\Lambda_3= \sin\vartheta \,\md\vartheta
\md\varphi$ for $\Lambda_3^i=(\sin\vartheta\cos\varphi,
\sin\vartheta\sin\varphi, \cos\vartheta)\in S^2$.

To reduce further to isotropy we have to describe in more detail the
space $\Uiso$. It can be written as $\Uiso=\bigcup_{c\in
  U(1)}\Theta_{c}$ where $\Theta_{c}$ are conjugacy classes in
$SU(2)^3$ with respect to the diagonal conjugation
$g(h_1,h_2,h_3)g^{-1}=(gh_1g^{-1},gh_2g^{-1},gh_3g^{-1})$ of $SU(2)$
on $SU(2)^3$. The $\Theta_{c}$ are labeled by an element
$c\in\R/(4\pi\Z)\cong U(1)$ and take the form
\begin{eqnarray*}
  \Theta_{c} & = & \{g(\exp (c\tau_1),\exp (c\tau_2),\exp (c\tau_3))
  g^{-1}: g\in SU(2)\}\\
  & = & \{(\exp (c\Lambda_1^i\tau_i),
  \exp (c\Lambda_2^i\tau_i), \exp (c\Lambda_3^i\tau_i)):\Lambda_I^i\in
  SO(3)\}\subset SU(2)^3.
\end{eqnarray*}
The components $\Lambda_I^i$ build a dreibein, which shows that
$\Theta_{c}$ for $c\not=0$ is homeomorphic to $SO(3)$. It is however
not a group, nor is $\Uiso$, which can e.g.\ be seen by multiplying
the elements $(\exp (c\tau_1),\exp (c\tau_2),\exp (c\tau_3))$ and
$(\exp (c\tau_2),\exp(-c\tau_1),\exp (c\tau_3))$, which are both
contained in $\Theta_{c}$.

An invariant normalized measure on the conjugacy class $\Theta_{c}$
is defined by
\begin{equation}
  \int_{\Theta_{c}}f(\Lambda)\,\md\mu_{\Theta_{c}}(\Lambda):=
  \int_{SU(2)}f(g(\exp (c\tau_1), \exp (c\tau_2),
  \exp (c\tau_3))g^{-1})\, \md\mu_H(g)\,.
\end{equation}
On the right hand side the element $(\exp (c\tau_1),\exp
(c\tau_2),\exp (c\tau_3))$ can be replaced by an arbitrary element of
$\Theta_{c}$; the measure $\md\mu_{\Theta_{c}}$ is independent of this
choice. The integration is only over one copy of $SU(2)$, because
$\Theta_{c}$ is defined as a conjugacy class with respect to the
diagonal conjugation of $SU(2)$ on $SU(2)^3$ (and not conjugation in
the group $SU(2)^3$).

Deleting the point $c=0$ we can represent $\Uiso$ as a fiber bundle
with fibers $U(1)\backslash\{1\}$ and base homeomorphic to $SO(3)$
represented by some $\Theta_{c}$. We can then build the product of
Haar measure $\md c$ on $U(1)$ weighted with the volume
$\Vol\Theta_{c}=(2\pi)^{-1}\sin^2(c/2)$ of the conjugacy classes and
the invariant measure on $\Theta_{c}$ to obtain a measure on $\Uiso$:
\[
  (2\pi)^{-1}\int_{U(1)}\int_{\Theta_{c}}f(\exp (c\Lambda_1),
  \exp (c\Lambda_2), \exp (c\Lambda_3))
  \md\mu_{\Theta_{c}}(\Lambda)\sin^2{\textstyle\frac{c}{2}}\, \md c.
\]
The point $c=0$ has no effect because it is of measure zero.

To make contact with reduced gauge invariant spin networks
consisting of a single edge we will now restrict ourselves to gauge
invariant functions $f$. Those functions do not depend on the dreibein
$\Lambda_I^i$, but only on the parameter $c$. This implies that any
such function can be written as $f(\exp (c\Lambda_1),
\exp (c\Lambda_2),\exp (c\Lambda_3))=F(\exp (c\Lambda_3))$ for some
function $F$ on $SU(2)$. Its integral over $\Uiso$ is
\begin{eqnarray*}
 & & (2\pi)^{-1}\int_{U(1)}\int_{\Theta_{c}}F(\exp (c\Lambda_3))
  \md\mu_{\Theta_{c}}(\Lambda)\sin^2{\textstyle\frac{c}{2}}\, \md c=\\ 
 & & \qquad(2\pi)^{-1}\int_{U(1)}\int_{SU(2)}
  F(g\exp (c\tau_3) g^{-1})\,\md\mu_H(g)\sin^2
  {\textstyle\frac{c}{2}}\,\md c
\end{eqnarray*}
in which, parameterizing $SU(2)$ with Euler angles
$g=\exp(\varphi\tau_3)\exp(\vartheta\tau_2)\exp(\psi\tau_3)$, the
$\psi$-integration is trivial. After performing this integration the
remaining part of the $SU(2)$-measure and the $U(1)$-measure recombine
to Haar measure on $SU(2)$ (this is Weyl's integral formula for the
group $SU(2)$ \cite{BroeckerDieck}) leading to the measure
\begin{equation}\label{Uisomeasure}
  \int_{\Uiso}f(h_1,h_2,h_3)\,\md\mu(h_1,h_2,h_3)=
  \int_{SU(2)}F(h_3)\,\md\mu_H(h_3)
\end{equation}
on $\Uiso$ for gauge invariant functions $f$. (Heuristically, the
measure $\md\mu_{\Theta_{c}}$ on $\Theta_{c}$ is
\[
  2\pi\sin^{-2}{\textstyle\frac{c}{2}}\delta(c'-c)\md\mu_H(g(c',\Lambda))
\]
and integrating the $\delta$-function replaces $c'$ by $c$.)

The measure just derived shows that the kinematical Hilbert space for
isotropic models can be represented as a space of functions on $SU(2)$
with the usual Haar measure. The spin networks with one edge
correspond to the character functions
\begin{equation}
  \chi_j(c)=\frac{\sin(j+\frac{1}{2})c}{\sin\frac{c}{2}}
  \quad, \quad j\in{\textstyle\frac{1}{2}}\N_0
\end{equation}
which are invariant under $c\mapsto-c$. But we have seen that these
functions do not suffice to generate all gauge invariant functions on
$\Uiso$. (To avoid misunderstanding, we note that the $\chi_j$
certainly span the space of class functions on $SU(2)$. However, the
gauge transformations on $\Uiso$ are not just conjugation on $SU(2)$,
but on a subspace of $SU(2)^3$. Therefore, there can be more gauge
invariant functions which do not reduce to class functions after
restricting to dependence on one edge only.) We are now going to
determine the remaining class of functions.

To that end we recall that any gauge invariant function on $\Uiso$ can
be written as a linear combination of functions $\tr[\exp
(m_1c\Lambda_1) \exp (m_2c\Lambda_2) \exp (m_3c\Lambda_3)]$ (up
to irrelevant factors of $\cos(c/2)$) with
some integers $m_I$. If some of the $m_I$ vanish the trace is
invariant under $c\mapsto -c$; otherwise it is easily evaluated to
\begin{eqnarray*}
  \tr[\exp (m_1c\Lambda_1) \exp (m_2c\Lambda_2) \exp
    (m_3c\Lambda_3)] & = & 2\cos({\textstyle\frac{1}{2}}m_1c)
    \cos({\textstyle\frac{1}{2}}m_2c)
    \cos({\textstyle\frac{1}{2}}m_3c)\\
  & &   -2\sin({\textstyle\frac{1}{2}}m_1c)
    \sin({\textstyle\frac{1}{2}}m_2c)
    \sin({\textstyle\frac{1}{2}}m_3c)\,.
\end{eqnarray*}
The first term is invariant under $c\mapsto -c$ and can thus be
expanded in the functions $\chi_j$. The second term, however, is
antisymmetric and provides new functions
\begin{eqnarray}
  & & \sin({\textstyle\frac{1}{2}}m_1c)
    \sin({\textstyle\frac{1}{2}}m_2c)
    \sin({\textstyle\frac{1}{2}}m_3c) = -{\textstyle\frac{1}{4}}
    \left[ \sin\left({\textstyle\frac{1}{2}}(m_1+m_2+m_3)c\right)
      \right.\label{expand}\\
  & &  +     \sin\left({\textstyle\frac{1}{2}}(m_1-m_2-m_3)c\right)
     + \left.\sin\left({\textstyle\frac{1}{2}}
       (-m_1-m_2+m_3)c\right)+
  \sin\left({\textstyle\frac{1}{2}}(-m_1+m_2-m_3)c\right)\right]\,. 
      \nonumber
\end{eqnarray}
All these functions with different sums $m_1+m_2+m_3$ are
independent provided we choose $m_I>0$.

\begin{lemma}\label{sine}
  All gauge invariant functions on $\Uiso$ can be
  generated by symmetric functions, which are spanned by the $\chi_j$,
  and the functions
  \[
    \sin (kc)\quad, \quad k\geq{\textstyle\frac{1}{2}}\,.
  \]
\end{lemma}

\begin{proof}
  As the discussion above shows, all independent functions are given
  by $\chi_j$ and the functions $\sin(m_1c/2)\sin(m_2c/2)\sin(m_3c/2)$
  for all different values of $m:=m_1+m_2+m_3$, $m_I>0$. In all cases
  $m$ can be decomposed as $m_1=m-2$, $m_2=m_3=1$. The expansion
  (\ref{expand}) then yields
  \[
    \sin\left({\textstyle\frac{1}{2}}(m-2)c\right)\sin^2
     {\textstyle\frac{c}{2}}=
    -{\textstyle\frac{1}{4}} \left[ \sin\left({\textstyle\frac{1}{2}}
      mc\right) -2
      \sin\left({\textstyle\frac{1}{2}}(m-2)c\right)+
      \sin\left({\textstyle\frac{1}{2}}(m-4)c\right) \right]\,.
  \]
  Induction over $m\geq3$ then shows that all independent antisymmetric
  functions are given by
  \[
    f_j:=\sin (jc)-\frac{j}{j-1}\sin((j-1)c)\quad,\quad
    j\geq{\textstyle\frac{3}{2}}\,.
  \]

  This set of functions can be simplified if we can generate the
  functions $\sin(c/2)$ and $\sin c$. This can indeed be achieved by
  using the sequence
  \[
    f_{\frac{3}{2}}+{\textstyle\frac{3}{5}}f_{\frac{5}{2}}+\cdots+
    {\textstyle\frac{3}{5}\cdot\frac{5}{7}
      \cdots\frac{M-2}{M}}f_{\frac{M}{2}}=
    -3\sin({\textstyle\frac{1}{2}}c)+ {\textstyle\frac{3}{M}}
    \sin({\textstyle\frac{1}{2}}Mc)
  \]
  for $M$ odd, and analogously for even $M$. In the Haar measure this
  sequence converges to $-3\sin(c/2)$ (the norm of $\sin(Mc/2)$ is
  independent of $M$), whereas for even $M$ we can obtain $\sin c$.

  Thus all antisymmetric functions are generated by $\sin (kc)$ with
  $k\geq\frac{1}{2}$.
\end{proof}

The kinematical Hilbert space is seen to be the linear span
$\langle\chi_j,\sin
kc:j\in\frac{1}{2}\N_0,k\in\frac{1}{2}\N\rangle$ completed in the
measure (\ref{Uisomeasure}).

\subsection{Derivative Operators}
\label{s:deriv}

Instead of the functions $\sin (kc)$ we will use functions which
appear more naturally when using derivative operators like
$\Lambda_3^i X^3_i$. As already noted, this operator maps a function
$\chi_j$ to a function being antisymmetric with respect to
$c\mapsto-c$. Writing the characters as
$\chi_j(c)=\tr\pi^{(j)}(g)=g^{(A_1}_{A_1}\cdots g^{A_{2j})}_{A_{2j}}$,
$g=\exp (c\Lambda_3)\in SU(2)$ we calculate
\begin{eqnarray*}
  \xi_j(c) & := & \Lambda_3^i X^3_i\,\chi_j(c)=
  2j\left.(\Lambda_3g)^{(A_1}_{A_1}\cdots
  g^{A_{2j})}_{A_{2j}}\right|_{g=\exp (c\Lambda_3)}\\
  & = & \frac{\md}{\md c}\chi_j(\exp (c\Lambda_3))=
  j\frac{\cos(j+\frac{1}{2})c}{\sin\frac{c}{2}}-
  {\textstyle\frac{1}{2}}\frac{\sin jc}{\sin^2\frac{c}{2}}
\end{eqnarray*}
noting that $\Lambda_3$ is a function on $SU(2)$ defined by $g=\exp
(c\Lambda_3)$.  Similarly, we can now justify the relation
$\Lambda_I^i J^3_i=0$ for $I\not=3$ used in simplifying the expression
(\ref{Jiso}): This operator leads to an insertion of
$\Lambda_I^i\tau_i$ into the trace of factors of $h_3=\exp
(c\Lambda_3)$ (with or without insertion of $\Lambda_3$) which
vanishes upon tracing.

The component $\Lambda_3^i X^3_i$ of the invariant vector field is
represented simply as derivative with respect to $c$. But this
derivative operator is, as already noted for $\Lambda_3^i X^3_i$, not
symmetric with respect to the measure (\ref{Uisomeasure}).  Therefore,
we now compute the adjoint of $\frac{\md}{\md c}$ and introduce new
functions related to $\xi_j$.

Due to
\begin{eqnarray*}
  \int^{4\pi}_0\overline{f(c)}g'(c)\sin^2{\textstyle\frac{c}{2}}\,\md c &
  = & -\int_0^{4\pi}\overline{f'(c)} g(c)
  \sin^2{\textstyle\frac{c}{2}}\,\md c- \int_0^{4\pi}\overline{f(c)}g(c)
  \sin{\textstyle\frac{c}{2}}\cos{\textstyle\frac{c}{2}}\, \md c\\
  & = & -\int_0^{4\pi}\left[\left(\frac{\md}{\md c}+
     \cot{\textstyle\frac{c}{2}}\right)
  \overline{f(c)}\right]g(c) \sin^2{\textstyle\frac{c}{2}}\, \md c
\end{eqnarray*}
we get
\begin{equation}
  \left(\frac{\md}{\md c}\right)^*=-\frac{\md}{\md c}-
   \cot{\textstyle\frac{c}{2}}
\end{equation}
as adjoint of $\frac{\md}{\md c}$ with respect to Haar measure.
Computing the commutator
\[
  \left[\frac{\md}{\md c},\left(\frac{\md}{\md c}\right)^*\right]=
  -\frac{\md}{\md c}\cot{\textstyle\frac{c}{2}}= 
   (2\sin^2{\textstyle\frac{c}{2}})^{-1}
\]
we see that $\frac{\md}{\md c}$ is not a normal operator.

More important for what follows will be the anti-selfadjoint
combination
\[
  {\textstyle\frac{1}{2}}\left(\frac{\md}{\md c}-\left(
  \frac{\md}{\md c}\right)^*\right)=
  \frac{\md}{\md c}+{\textstyle\frac{1}{2}}\cot{\textstyle\frac{c}{2}}\,.
\]
By means of this derivative operator we define our final class of
functions
\begin{equation}
  \zeta_j(c):=(j+{\textstyle\frac{1}{2}})^{-1}
  \left(\frac{\md}{\md c}+{\textstyle\frac{1}{2}}
    \cot{\textstyle\frac{c}{2}}\right) \chi_j(c)=
  \frac{\cos(j+\frac{1}{2})c}{\sin\frac{c}{2}}\quad,\quad
  j\in{\textstyle\frac{1}{2}}\N_0
\end{equation}
and
\[
  \zeta_{-\frac{1}{2}}:=(\sqrt{2}\sin{\textstyle\frac{c}{2}})^{-1}
\]
which are antisymmetric in $c$. (The singularity in $c=0$ is
compensated in the Haar measure.) We now take a closer look on the
action of $\frac{\md}{\md c}+\frac{1}{2}\cot\frac{c}{2}$. Acting on
$\chi_j$ yields by definition
\begin{equation}\label{dchi}
  \left(\frac{\md}{\md c}+{\textstyle\frac{1}{2}}\cot{\textstyle\frac{c}{2}}
  \right)\chi_j(c)=
  \left(j+{\textstyle\frac{1}{2}}\right)\zeta_j(c)
\end{equation}
and acting on $\zeta_j$
\begin{equation}\label{dzeta}
  \left(\frac{\md}{\md c}+{\textstyle\frac{1}{2}}\cot{\textstyle\frac{c}{2}}
  \right)\zeta_j(c)=
  -\left(j+{\textstyle\frac{1}{2}}\right)\chi_j(c).
\end{equation}

The functions $\zeta_j$ are easily seen to be orthonormal, which is
also true for the functions $\chi_j$.  Furthermore, each $\zeta_{j_1}$
is orthogonal to any $\chi_{j_2}$ because their product is
antisymmetric in $c$. Thus, the set
\begin{equation}
  \left\{\chi_{j_1},\zeta_{j_2}: j_1\in{\textstyle\frac{1}{2}}\N_0,
    j_2\in{\textstyle\frac{1}{2}}\N_0\cup
    \{-{\textstyle\frac{1}{2}}\}\right\}
\end{equation}
is an orthonormal basis of functions in $c$ with respect to the
measure $(2\pi)^{-1}\sin^2\frac{c}{2}$.

We can build another selfadjoint operator from $\frac{\md}{\md c}$,
namely the selfadjoint quadratic differential operator
\begin{equation}
  \left(\frac{\md}{\md c}\right)^*\frac{\md}{\md c}=-\frac{\md^2}{\md c^2}-
  \cot{\textstyle\frac{c}{2}}\,\frac{\md}{\md c}\,,
\end{equation}
which is the radial component of the Laplace operator on $SU(2)$, i.e.\
the part independent of $\vartheta,\varphi$ being the only
non-vanishing contribution when acting on gauge (conjugation) invariant
functions. Indeed, it is straightforward to show, that
\[
  \left(\frac{\md}{\md c}\right)^*\frac{\md}{\md c}\,
   \chi_j(c)=j(j+1)\chi_j(c)
\]
for the class functions $\chi_j$. Our additional functions $\zeta_j$
are also eigenfunctions with the same eigenvalues:
\[
  \left(\frac{\md}{\md c}\right)^*\frac{\md}{\md c}\,\zeta_j(c)=
   j(j+1)\,\zeta_j(c)\,,
\]
which follows from the definition of $\zeta_j$ and
\[
  \left[\left(\frac{\md}{\md c}\right)^*\frac{\md}{\md c},
   \frac{\md}{\md c}\right]=
  \left[\left(\frac{\md}{\md c}\right)^*\frac{\md}{\md c},
   \left(\frac{\md}{\md c}\right)^*\right]\,.
\]

However, neither the functions $\xi_j$ defined earlier nor the
functions $\sin jc$ are eigenfunctions of $\left(\frac{\md}{\md
    c}\right)^*\frac{\md}{\md c}$. We have, for instance,
\[
  \left(\frac{\md}{\md c}\right)^*\frac{\md}{\md c}\sin jc= 
   j(j+1)\sin jc-
  j\frac{\cos(j-\frac{1}{2})c}{\sin\frac{c}{2}}= j(j+1)\sin
  jc- j\zeta_{j-1}(c)
\]
showing that the function $\zeta_j$ can be obtained by acting with
$\left(\frac{\md}{\md c}\right)^*\frac{\md}{\md c}$ on $\sin
((j+1)c)$. Vice versa, we can reobtain the function $\sin jc$ by using
the equations
\begin{equation}
  \sin (jc)={\textstyle\frac{1}{2}}\left(\zeta_{j-1}(c)-
  \zeta_j(c)\right)\quad, \quad j>{\textstyle\frac{1}{2}}
\end{equation}
and
\[
  \sin{\textstyle\frac{c}{2}}={\textstyle\frac{1}{2}}
  \left(\sqrt{2}\,\zeta_{-\frac{1}{2}}-\zeta_{\frac{1}{2}}\right)
\]
which follow from adding the two equations
\[
  \sin{\textstyle\frac{c}{2}}\sin (jc)=\cos\left(\left(
    j-{\textstyle\frac{1}{2}}\right)c\right)-
  \cos{\textstyle\frac{c}{2}}\cos (jc)= \cos{\textstyle\frac{c}{2}}\cos (jc)-
  \cos\left(\left(j+{\textstyle\frac{1}{2}}\right)c\right)
\]
and dividing by $\sin (c/2)$.

We now arrived at our final set of generating functions:

\begin{theo}
  The set
  \begin{equation}
   \left\{\chi_j,\zeta_k:j\in
     {\textstyle\frac{1}{2}}\N_0,k\in
     {\textstyle\frac{1}{2}}\N_0 \cup
     \left\{-{\textstyle\frac{1}{2}}\right\} \right\} 
  \end{equation}
  forms an orthonormal basis of the kinematical Hilbert space of
  isotropic models $\Haux=L^2(\Uiso,\md\mu_H)$.
\end{theo}

\begin{proof}
  According to Lemma \ref{sine} all antisymmetric functions on $\Uiso$
  can be generated by the functions $\sin jc$ for $j\geq\frac{1}{2}$.
  With the preceding equations we see that this set of functions is
  equivalent to the set $\{\zeta_j: j\geq-\frac{1}{2}\}$.  That all
  functions contained in the set of generating functions are
  orthonormal has already been shown above.
\end{proof}

\subsection{Isotropic Volume}
\label{s:isovol}

Finally, we have to translate the derivative operators $\frac{\md}{\md
  c}$ back to invariant vector field operators $X_i^3$. We were lead
to $\frac{\md}{\md c}$ by studying the action of $\Lambda_3^i X^3_i$
on spin network functions, which is not selfadjoint. The selfadjoint
operator $\frac{1}{2}(\Lambda_3^i J^3_i+(\Lambda_3^i J^3_i)^*)$ is now
identified with the derivative operator
\[
  {\textstyle\frac{i}{2}}\left(\frac{\md}{\md c}-
   \left(\frac{\md}{\md c}\right)^*\right)=
  i\frac{\md}{\md c}+{\textstyle\frac{i}{2}}\cot{\textstyle\frac{c}{2}}\,,
\]
which we regard as quantization of the classical real phase space
function $\Lambda_3^i E^3_i$. Its action on the quantum states
$\chi_j$ and $\zeta_j$ for $j\geq0$ is (using equations (\ref{dchi}) and
(\ref{dzeta}))
\begin{eqnarray}
 \widehat{\Lambda_3^i E^3_i}\,\chi_j & = &
 i\left(j+{\textstyle\frac{1}{2}}\right)\zeta_j\,,\\
 \widehat{\Lambda_3^i E^3_i}\,\zeta_j & = &
 -i\left(j+{\textstyle\frac{1}{2}}\right)\chi_j\,,
\end{eqnarray}
whereas it annihilates $\zeta_{-\frac{1}{2}}$.

For the volume operator we need the spectrum of
\[
  \left|\widehat{\Lambda_3^i E^3_i}\right|:= \sqrt{\widehat{\Lambda_3^i
    E^3_i}^2}\:,
\]
which can be read off from the previous equations. It has the
twofold degenerate eigenvalues $j+\frac{1}{2}$ for $j\geq0$ with
eigenfunctions $\chi_j$ and $\zeta_j$ and the non-degenerate eigenvalue
$0$ with eigenfunction $\zeta_{-\frac{1}{2}}$.

The operator $(J^3)^2$, which commutes with $\widehat{\Lambda_3^i
  E^3_i}$, is to be identified with the quadratic derivative
$\left(\frac{\md}{\md c}\right)^*\frac{\md}{\md c}$ with eigenvalues
$j(j+1)$ to the same eigenfunctions as above.

With these ingredients we can now quantize the volume (\ref{volisotropic})
using equations (\ref{Jiso}) and (\ref{volbianchiquant}) as
\begin{equation}
  \hat{V}=V_0 (\iota' l_P^2)^{\frac{3}{2}}
   \sqrt{\left|\widehat{\Lambda_3^i E^3_i}\right|(J^3)^2}\:.
\end{equation}
The spectrum is then easily obtained using the information presented
above as
\begin{equation}\label{isospec}
  \left\{V_0(\iota'l_P^2)^{\frac{3}{2}}
    \sqrt{j\left(j+{\textstyle\frac{1}{2}}\right)(j+1)}:
    j\in{\textstyle\frac{1}{2}}\N_0\right\}
\end{equation}
which is twofold degenerate for $j>0$, whereas $j=0$ is triply
degenerate.

\subsection{Remarks on LRS Models}

The general features of solving the Higgs constraint in the spin
network context to arrive at the kinematical Hilbert space are
illustrated by the example of isotropic models, which was considered
in detail above. E.g., one has to determine the quantum states
with its possible insertions and to carry over the spin network
techniques. Conceptually, the situation for LRS models is the same,
but it is complicated by the appearance of two edges and, in
connection, the dependence of gauge invariant states on two variables.
We showed also in this case the necessity of insertions. All other
steps are to be done in analogy to isotropic models. We will not
present them here because they do not bring in anything new.

\section{Consequences and Discussion}
\label{s:applic}

In this final section we comment on some applications of the material
contained in the present paper.

\subsection{Quantum Symmetry Reduction}

The isotropic models considered in detail in Section~\ref{s:iso}
provide the first example of a symmetry reduction with nontrivial
Higgs sector and non-empty Higgs constraint being carried out
completely along the lines of quantum symmetry reduction
\cite{SymmRed}. They show a concrete illustration of how to solve the
Higgs constraint in quantum theory by using the geometrical Higgs
constraint (\ref{Higgs}). Furthermore, the need for insertions and
their interpretation as remnants of vertex contractors has shown
up. The treatment proved that spin network techniques can be adapted
to solution spaces of the Higgs constraint. An essential ingredient
was to relax the partial gauge fixing, thereby restoring the full
$SU(2)$-gauge invariance.

More complicated models are provided by locally rotationally symmetric
systems which lead to spin networks with an axial and a transversal
edge, and which can be treated along the same lines. The transversal
edge represents the information contained in the two edges which are
equivalent upon solving the Higgs constraint. This is similar to the
spherically symmetric sector of loop quantum gravity
\cite{SymmRed,AreaOp}, where instead of the axial edge (representing a
point holonomy and not a real edge) we have a radial manifold on which
Higgs vertices are lined up. These vertices also contain one edge (in
an auxiliary manifold) which is obtained after reducing two
transversal edges when solving the Higgs constraint. Therefore, LRS
models are good toy models for determining the structure of
spherically symmetric Higgs vertices.  This was our main motivation
for studying cosmological models, because single Higgs vertices can
here be investigated on their own.

\subsection{Level Splitting}

In the volume spectra we can see a phenomenon first observed in case
of the area spectrum in reference \cite{AreaOp}. Starting from the
full spectrum of reference \cite{Vol2}, which is, however, not known
explicitly, we obtain only a subset of this spectrum after reducing to
homogeneous geometries. This is a consequence of the fact that there
is only one point $x_0$ in the reduced manifold $B$, and therefore
only one vertex. The vertex sum in the full volume spectrum then
disappears, and the spectrum is reduced because the eigenvalues are
in general irrational. We then can enhance the symmetry further to LRS
and, finally, isotropic models, which have the simple volume spectrum
(\ref{isospec}).

Vice versa, starting from isotropic models the symmetry can be broken
in steps to finally obtain an arbitrary anisotropic, inhomogeneous
geometry. In each step the broken symmetry leads to a splitting of
eigenvalues of the volume operator leading from the spectrum
(\ref{isospec}) to the full spectrum. Note that we can only compare
the eigenvalues, not the degeneracies, because reduced models are
represented in different Hilbert spaces which are not subspaces of the
full Hilbert space. Alternatively, their states can be described by
distributional states of the full theory as described in reference
\cite{SymmRed}. In particular, we cannot determine which eigenvalues
of the full theory are related by this level splitting to a particular
eigenvalue in the spectrum (\ref{isospec}). The degeneracy of two for
$j>0$ in this spectrum has nothing to do with such a degeneracy
expected from level splitting.

Another feature of the high symmetry of isotropic models is that we
can explicitly calculate the complete volume spectrum, a task which
would be hopeless in the full theory.

\subsection{Weaves}

As made explicit by a spin network, the quantum nature of gravity
breaks explicitly any continuous space symmetry. A nontrivial spin
network (or a finite linear combination) cannot be invariant with
respect to a transitive symmetry group.  Therefore, our homogeneous or
isotropic states are, when not regarded only as states of a reduced
toy model, idealized states comparable to plane waves in quantum
mechanics.  Accordingly, they are represented as generalized states of
the full theory \cite{SymmRed}, i.e.\ as elements of the topological
dual $\Phi'$ of the space $\Phi$ of cylindrical functions. But using a
nuclear topology of $\Phi$, this space is dense in its dual in the
weak topology \cite{Gelfand4}.  Therefore, any distributional state
can be approximated weakly by certain combinations of spin network
states. Although such an approximation may be very complicated to
construct explicitly, this provides a simple existence proof for
$S$-weave states. We define here those states as states which
approximate a given generalized state being symmetric with respect to
the symmetry group $S$. For instance, we can build isotropic
$S$-weaves by approximating the states found in Section~\ref{s:iso},
regarded as distributional states of the full theory using the map
$\sigma_{[\lambda]}$ of reference \cite{SymmRed}.

We denote them as $S$-weaves to point out that they are not
necessarily equivalent to the weaves defined in reference
\cite{Weaves}. There states were defined as weaves which approximate a
given classical metric at large distances as compared to the Planck
scale. Such a geometrical condition is not contained in our definition
of $S$-weaves, and the meaning of approximation is different in both
cases. Note, however, that our definition and the existence proof are
not trivial, because it is not obvious how to construct finite linear
combinations of spin networks whose inner product with any other spin
network is approximately independent of its position (already before
solving the diffeomorphism constraint). But a connection between both
concepts of weave states exists. For suppose that we solved the
Hamiltonian constraint of the isotropic model associated with Bianchi
I \cite{cosmoIII}, and we found a distinguished solution representing
the unique classical solution, namely Minkowski space-time, we can
approximate this solution by an $S$-weave for its associated
distributional state. This $S$-weave is then expected to contain,
besides its approximate isotropy, geometrical information
approximating the Euclidean metric of space.

However, the $S$-weaves are only approximately symmetric. They
manifestly break the symmetry, and therefore applying
the volume operator will not lead to a spectrum of the simple form
(\ref{isospec}), but rather of the form of the complicated spectrum in
the full theory. Consequences of this fact will now be considered in a
final subsection.

\subsection{Cosmology}

In cosmology one usually reduces a theory of gravity classically to
homogeneous metrics reducing the degrees of freedom to finitely
many. Dynamics is then encoded in the Wheeler--DeWitt equation which is
a hyperbolic differential equation with respect to the scale factor
\cite{DeWitt}. To account for fluctuations which are necessary for
structure formation, however, one has to disturb the homogeneous
geometries and can treat the ensuing inhomogeneities as perturbations
\cite{Halliwell}. In an appropriate neighborhood of homogeneous models
the Wheeler--DeWitt equation will remain hyperbolic in one variable
\cite{Giulini}. 

But we have seen that in a theory using quantized geometries even
small perturbations have drastic effects: We can approximate a
symmetric state arbitrarily precisely by $S$-weaves, which represent
slightly perturbed symmetric metrics, but the volume spectrum of these
states will never reach the simple spectrum (\ref{isospec}) for
isotropy. This is so because $S$-weaves are ordinary, however
complicated, spin network states which, when chosen as volume
eigenstates, correspond to a volume eigenvalue with a large vertex
sum. Thus, we see that homogeneous metrics are very special in a
quantum theory, and in view of the present paper results obtained
with minisuperspace models are unlikely to be reproduced in a full
quantum theory of gravity. Even minor perturbations break the special
features of symmetric states, e.g.\ concerning the volume spectrum, to
full extent.

Of course, up to now our discussion remained at the kinematical level,
and the role of these kinematical properties after solving the
Hamiltonian constraint has not been investigated yet. However,
for dynamics the volume operator plays an important role, too, for it
appears quite naturally in the quantized Hamiltonian constraint
\cite{QSDI,cosmoIII}. Therefore, the kinematical volume spectrum is
significant for dynamics, and its features associated with symmetry
reduction should be expected to have a great impact on dynamics.

Finally, we note that the models discussed here may provide new
insights into the issue of the Hamiltonian constraint. Already the
simple geometries of reduced spin networks (for instance, only one
edge and an insertion for isotropic models) simplify its action
considerably. Maybe more important is the fact that the volume
spectrum is simplified, and even completely known in case of isotropy,
which facilitates determining the matrix elements of the constraint.

\section*{Acknowledgements}

The author is grateful to H.\ Kastrup for comments on the paper. He
also thanks the DFG-Graduierten-Kolleg ``Starke und
elektroschwache Wechselwirkung bei hohen Energien'' for a PhD
fellowship.

\end{document}